\documentclass[conference]{IEEEtran}
\usepackage{amsmath,graphicx}
\usepackage{nonfloat}

\usepackage{multirow}
\usepackage{graphicx}
\usepackage{subfig}
\title{Wisdom of the Crowd: Incorporating Social Influence in Recommendation Models\footnotemark}

\author{
\IEEEauthorblockN{Shang Shang\IEEEauthorrefmark{1}, Pan
Hui\IEEEauthorrefmark{2}, Sanjeev R. Kulkarni\IEEEauthorrefmark{1}, and Paul W. Cuff\IEEEauthorrefmark{1}}
\IEEEauthorblockA{\IEEEauthorrefmark{1}Department of Electrical Engineering,
Princeton University, Princeton NJ, 08540, U.S.A.
} \IEEEauthorblockA{\IEEEauthorrefmark{2}Deutsche Telekom
Laboratories, Ernst-Reuter-Platz 7, 10587 Berlin, Germany\\
\IEEEauthorrefmark{1}\{sshang, kulkarni, cuff\}@princeton.edu, \IEEEauthorrefmark{2}pan.hui@telekom.de}
}

\begin{document}
\maketitle
\let\thefootnote\relax\footnote{This research was supported in part by the Center for Science of
Information (CSoI), an NSF Science and Technology Center, under grant
agreement CCF-0939370, by the U.S. Army Research Office under grant
number W911NF-07-1-0185, and by a research grant from Deutsche Telekom
AG.}
\begin{abstract}

Recommendation systems have received considerable attention recently. However, most research has been focused on improving the performance of collaborative filtering (CF) techniques. Social networks, indispensably,  provide us extra information on people's preferences, and should be considered and deployed to improve the quality of recommendations. 

In this paper, we propose two recommendation models, for individuals and for groups respectively, based on \emph{social contagion} and \emph{social influence network theory}. In the recommendation model for individuals, we improve the result of collaborative filtering prediction with \emph{social contagion} outcome, which simulates the result of information cascade in the decision-making process. In the recommendation model for groups, we apply \emph{social influence network theory} to take interpersonal influence into account to form a settled pattern of disagreement, and then aggregate opinions of group members. By introducing the concept of susceptibility and interpersonal influence, the settled rating results are flexible, and inclined to members whose ratings are ``essential''.  
%{\bf Keywords:} recommendations systems, social influence

\end{abstract}

%These recommendation models formulate how people's preferences are affected by social influence, thus give us some incentives on general marketing strategies for new product release. In the future, we want to further our research by making recommendations to target items to individuals and group users.    

\begin{keywords}
recommendation model, social influence, collaborative filtering
\end{keywords}

\section{Introduction}
With the rise of e-commerce, recommendation systems have been studied intensively in the context of collaborative filtering (CF) techniques \cite{Su}. Early generations of recommendation system have already been commercialized and have achieved great success. Recommendation systems serve as an important component of online retail and Video on Demand (VoD) such as Amazon and Netflix \cite{Minh}. Recommenders give customized recommendations to online users on books, movies, and commodities according to their previous preference data. 
%\footnote{\url{www.netflix.com}} \footnote{\url{www.amazon.com}}
In order to improve the quality of recommendations, work has been done mostly based on improving collaborative filtering techniques. Most recommendation systems based on collaborative filtering assume that users are independent, which ignores the role of social influence in people's buying decisions. Problems such as ``data sparsity'', ``cold start'', ``shilling attack'' still challenge the design of recommendation systems \cite{Su}\cite{He}.  Social networks and social influence, on the other side, can provide us extra information on users' preferences,  but have received less attention. This may be partially due to the unavailability of large-scale datasets to analyze in the past. Recently, the emergence of Online Social Networks (OSN) provides us an opportunity to reconsider the structure and effects of social networks so as to improve recommendation results \cite{He}\cite{Palau}\cite{Zheng}. He et al. proposed a social network-based recommendation system (SNRS) in \cite{He}, which is a probabilistic model for personalized recommendations.  SNRS is based on \textit{homophily} among friends but did not take social influence into consideration and did not provide a solution for recommendation to groups. Traditionally, recommendation systems are designed to recommend items to individual users. However, there are some buying-decisions/activities done by a group of users such as going to a restaurant or watching a movie where group recommendations are desired. Surprisingly, recommendation to groups is a nontrivial extension of recommendation to individuals. Unlike individual's buying decisions which are a personal choice (the user may be influenced by others' opinions but there will be no compromise in the final decision), preference of a group reflects not only each individual's subjective taste but also the knowledge of other group members' opinions. Most of the current group recommendation systems such as POLYLENS \cite{O'Connor} and G.A.I.N. \cite{Pizzutilo} focus on different ways to aggregate preference. However, the problem of how to make recommendations for a group with the consideration of interpersonal influence among group members rather than viewing each individual's preference separately is often ignored.

In this paper, we propose two mathematical models of recommendation systems, for individuals and for groups respectively, based on \emph{social contagion} and \emph{social influence network theory}. In the past few years, study on \emph{social contagion} finds applications in viral marketing, which uses pre-existing social networks to promote new products or brands to market, without the knowledge of users' preference records \cite{Cosley}. Given the information of users' social networks and preference data, we use a modified linear threshold contagion model to simulate the social influence by word of mouth, and add this effect to collaborative filtering results in order to provide more effective recommendations to individuals. In the model of recommendations to groups, in order to consider people's ``compromised disagreement'' among group members,  we introduce social influence under sufficient interpersonal communication to give recommendations that reflect more than personal taste.

In a typical setting, there is a list of $m$ users $\mathcal{U} = \{u_1, u_2, ..., u_m\}$ and a list of $n$ items $\mathcal{I} = \{i_1, i_2,..., i_n\}$. Each user $u_j$ has a list of items $I_{u_j}$, which the user has rated or from which the user's preferences can be inferred. The ratings can either be explicit, for example, on a 1-5 scale as in Netflix, or implicit such as purchases or clicks. These data form a $m \times n$ rating matrix. $\mathcal{G} = (\mathcal{U},\mathcal{E})$ is a social network, represented by an undirected graph, where $\mathcal{U}$ is a set of nodes and $\mathcal{E}$ is a set of edges.  For all $ u, v \in \mathcal{U}$, $(u,v)\in \mathcal{E}$ if $u, v$ are ``friends'', in which case $v\in N(u)$ and $u\in N(v)$, where $N(u)$ is the set of neighbors of $u$. We want to make recommendations for a target user or a group of users given the above information. 

The rest of this paper is organized as follows. We discuss related work in Section \ref{sec: related work}. We propose our social influential recommendation models for individuals and for groups in Section \ref{sec: individuals} and Section \ref{sec: groups}, followed by conclusions and future work in Section \ref{sec: conclusions}.

\section{Related Work}
\label{sec: related work}
A few recent papers used information of users' social networks as a component in their recommendation models. The work in \cite{Aranda} used friendship matrix to modify the probabilistic matrix factorization proposed in \cite{Minh}. In \cite{Debnath}, Debnath et al. constructed a social network graph with items as nodes, which represents human judgement of similarity between items aggregated over a large population of users to estimate feature weights for the content-based recommendations.  The work in \cite{Palau} studied how collaboration should be done in a recommendation system based on a network of ``personal agents'' . The most closely related work is found in \cite{Zheng} and \cite{He}. The work in \cite{Zheng} proposed to reduce computational cost by limiting similar users to target user's immediate friends. The work in \cite{He} considered the effects of homophily among friends in a recommendation system and built a probabilistic model to describe this homophily effect.  In \cite{He}, the author assumed that item attributes, user attributes and ratings of immediate friends are independent of each other and that for each pair of immediate friends, their ratings on the same item are identical but with an error term following the distribution of the histogram of previous rating differences. There are, to the best knowledge of the authors', no previous work considering social influence, which plays an important role in people's buying decisions. We use a \emph{social contagion} model to add the effects of information cascade to the target user in the buying decision, in order to improve the result of collaborative filtering prediction. 

In addition to recommendation for individuals, there are some circumstances that recommendation for a group of users are desired. There are mainly three categories of recommendation systems for groups \cite{Jameson}: (1) merging sets of recommendations, e.g. taking intersection among all members' top-$k$ preference; (2) aggregation of individuals' ratings for particular items; (3) construction of group preference models. POLYLENS  \cite{O'Connor}, a system to recommend movies to a group of users, merged the results of individual recommendations. While Yu et al. \cite{Yu} introduced a group preference model which merged individual profiles, using a minimization of total Dalal's distance, and took this virtual profile as a target user for recommendations. In this paper, we utilize \emph{social influence network theory} \cite{Friedkin} as a tool to model the opinion formation of a group of users under interpersonal influences within the group.

\section{Recommendations for individuals}\label{sec: individuals}
Traditional collaborative filtering systems suppose that users are all independent. However, in reality, social influence plays a crucial role in people's buying decisions. For example,  a user, Jack, is browsing on Amazon to choose a leisure book to read, with nothing particular in mind. In the list of new collections, Jack vaguely remembers that his friend Lisa mentioned book A in the list. However, it is a romance book, which is not his type. His eyes then are caught by two other books B and C. B, a thriller, was strongly recommended by his friends Henry and James, and C, though, he has never heard of it, also looks interesting. Finally, Jack puts C in wish list and decides to buy book B first. If we look closely at this scenario, and think about how we make our decisions everyday, we can see that what drives us to buying decisions are not only our own preferences but also some informative comments made by friends. In addition, studies show that two persons connected via a social relationship tend to have similar tastes, which is known as ``\emph{homophily principle}'' \cite{He}. Social influence and social networks thus can help us predict a target user's preferences and decisions. In this section, we will first briefly introduce a prevalent memory-based collaborative filtering algorithm as a baseline algorithm, and then describe our social influential recommendation model for individual users in detail. 

\subsection{Collaborative Filtering}

Collaborative filtering (CF) is one of the most successful approaches to build a recommendation system. It uses the known preferences of users to make recommendations or predictions to the target user \cite{Su}. 

\subsubsection{Similarity Computation}
There are a variety of similarity measures. A generally adopted one is called \emph{Pearson Correlation} which measures the extent to which two variables linearly relate with each other \cite{Resnick}. For user-based algorithm, the \emph{Pearson Correlation} between user $u$ and $v$ is

\begin{equation}
\label{}
\displaystyle
w_{u,v} = \frac{\sum_{i \in I}(r_{u,i} -\bar{r}_u)(r_{v,i} -\bar{r}_v)}{\sqrt{\sum_{i \in I}(r_{u,i}-\bar{r}_u)^2}\sqrt{\sum_{i \in I}(r_{v,i}-\bar{r}_v)^2}},
\end{equation}
where $i \in I$ is the item rated by both users $u$ and $v$, $r_{u,i}$ is the rating of user $u$ on item $i$, and $\bar{r}_u$ is the average rating of user $u$ in co-rating set $I$.

\subsubsection{Prediction Computation}

Prediction Computation is the most important step in a collaborative filtering system \cite{Su}. We can use the weighted sum to predict the rating $P_{u,i}^{cf}$ for target user $u$ on a certain item $i$ as proposed in \cite{Resnick}:

\begin{equation}
\label{}
\displaystyle
P_{u,i}^{cf} = \bar{r}_u + \frac{\sum_{v \in \mathcal{U}}(r_{v,i} - \bar{r}_v) \cdot w_{u,v}}{\sum_{v \in \mathcal{U}}|w_{u,v}|}. 
\end{equation}

Recommenders based on collaborative filtering then refer to this prediction to provide top-$k$ recommendations to the user or simply display the personalized predicted rating of each item. Unfortunately, they ignore the dependence and influence in social networks in users preferences. \emph{Social Contagion Model} provides us an alternative solution to deal with this problem by taking these effects into consideration. 

\subsection{Social Contagion Model}
The flow of information through a social network can be thought of as unfolding with the dynamics of an epidemic \cite{Kleinberg}.  This information from social relationships has potential influence in people's final buying decision. We use a  \emph{linear threshold social contagion model} \cite{Kempe}\cite{Kleinberg} to make recommendations with respect to this natural social contagion and user's subjective taste. We will start with a simple case that users use ``like'' and ``dislike'' to describe their personal preferences and explain the model in the perspective of game theory. We will then discuss the multiple-scale rating case in Section \ref{subsec:general}.

\subsubsection{A Simple Binary Rating Case}\label{subsec:simple}
In a binary rating system, we assume that users have three possible states namely \emph{like}, \emph{dislike} and \emph{inactive}. We use ``1'', ``-1'' and ``inactive-0'' to label them respectively. A node $v$ is influenced by each neighbor $w$ according to a weight/influence factor, $b_{v,w} \in [0,1]$ such that $\sum_{w \in N(v)}b_{v,w} = 1$. Intuitively, $b_{v,w}$ is related to trust and interpersonal communication frequency. If we lack such knowledge, we can randomly partition the unit influence among neighbors of a user, which in fact simulates the randomness of information cascade. The process proceeds as follows: each node $v$ chooses a threshold $\theta _{v,i}$ for item $i$  from the interval $[0,1]$, representing the weighted fraction of $v$'s neighbors that must become active (either  \emph{like} or \emph{dislike} in order for $v$ to become active to state \emph{like} or \emph{dislike} on item $i$). Since $\theta_{v,i}$ indicates the latent tendency of nodes to adopt the opinion, we naturally associate it with $v$'s CF prediction on item $i$. E.g. we can set it as in Equation \ref{eq:theta }.  Two other classes of approaches are setting all thresholds uniformly at random from the interval [0,1] or at a known value like 1/2. Given a threshold, and an initial set of active nodes $A_i$ (users with non-empty ratings on a certain item $i$), this progressive diffusion process proceeds deterministically.

Time operates in discrete steps $t = 1,2,3, \ldots.$ At a given time $t$, any inactive node $v$ becomes active if its fraction of active neighbors exceeds its threshold:

\begin{equation}
\label{eq:threshold}
\left|\sum_{w\in N(v)}b_{vw} \cdot S_{w,i}\right| \ge \theta_{v,i},
\end{equation}
where $S_{w,i}$ is the state of node $w$ on item $i$, and $\theta_{v,i}$ is the influential threshold.

\begin{equation}
\displaystyle
\label{eq:theta }
\theta_{v,i}  = \left\{\begin{array}{ll} P_{v,i}^{cf}, \quad \textrm{if }P_{v,i}^{cf}\cdot\left(\sum_{w\in N(v)}b_{vw} \cdot S_{w,i}\right) < 0  \\ \min\{P_{v,i}^{cf}, 1- P_{v,i}^{cf} \}, \quad \textrm{otherwise}\end{array}\right..
\end{equation}

The intuition in Equations \ref{eq:threshold} and \ref{eq:theta } is that if receiving too many controversial opinions from neighbors, the node will fail to be activated to either state. Equation \ref{eq:theta } states that similar opinions to users' expectations are easier and opposite opinions are harder to be taken. 

The state of newly activated node $v$ under social influence is decided by, 
\begin{equation}
\label{ }
S_{v,i} = \textrm{sign}\left(\sum_{w\in N(v)}b_{vw} \cdot S_{w,i}\right)
\end{equation}

This in turn may cause other nodes to become active in subsequent time steps, leading to potentially cascading adoption behaviors of ``like'' or ``dislike''. The process runs until no more activations are possible. 

For the target user $u$,
\begin{equation}
\label{ }
P_{u,i}^{si} = \left\{\begin{array}{ll} S_{u,i} & \textrm{if }u \textrm{ is activated}  \\ \textrm{inactive-}0 & \textrm{otherwise}\end{array}\right..
\end{equation}

\begin{figure*}[!hbtp] % use float package if you want it here
\centering
\subfloat[Threshold = 0.1. ]{\label{fig:1}\includegraphics[width=0.33\textwidth]{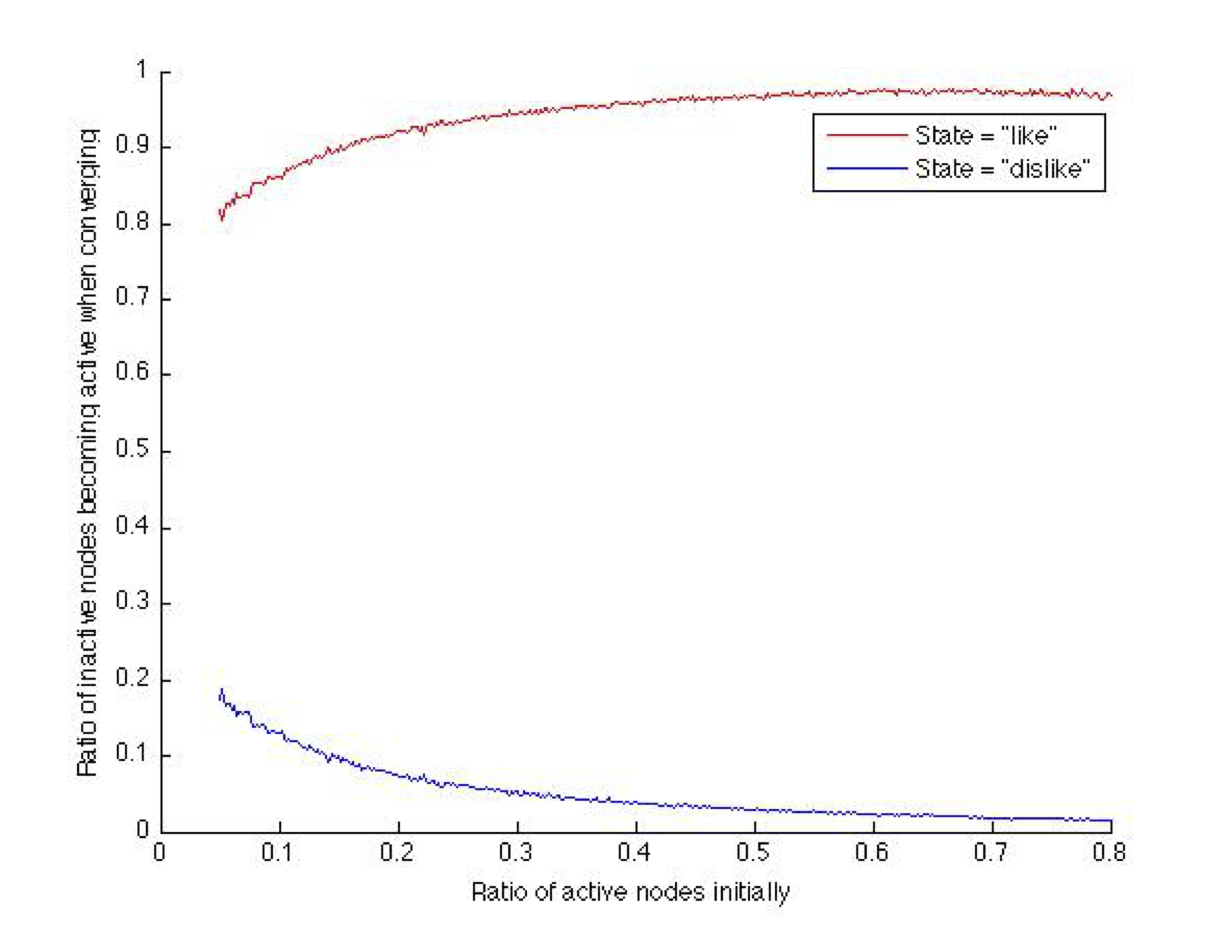}}                
\subfloat[Threshold = 0.5.]{\label{fig:2}\includegraphics[width=0.33\textwidth]{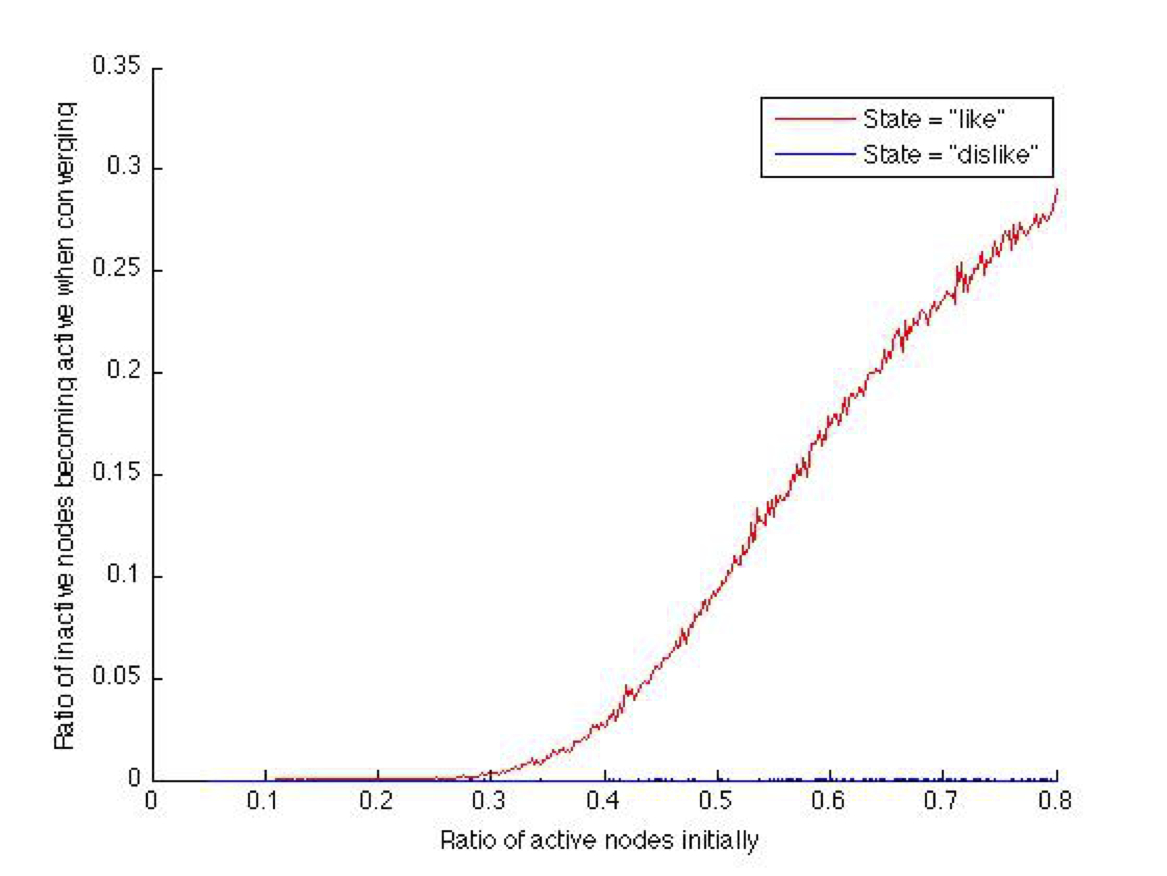}}
\subfloat[Threshold uniformly drawn from 0.05 to 0.8.]{\label{fig:3}\includegraphics[width=0.33\textwidth]{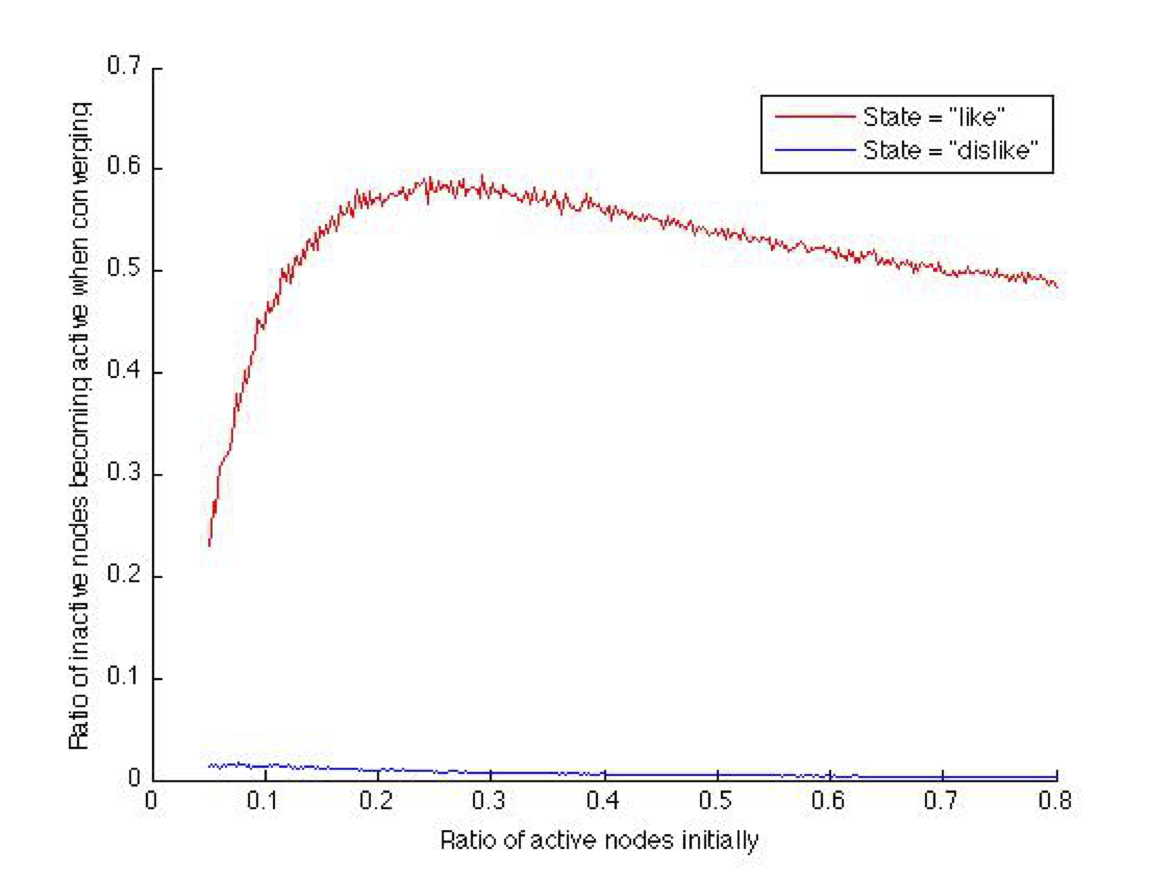}}\\
\subfloat[Threshold = 0.1.]{\label{fig:4}\includegraphics[width=0.33\textwidth]{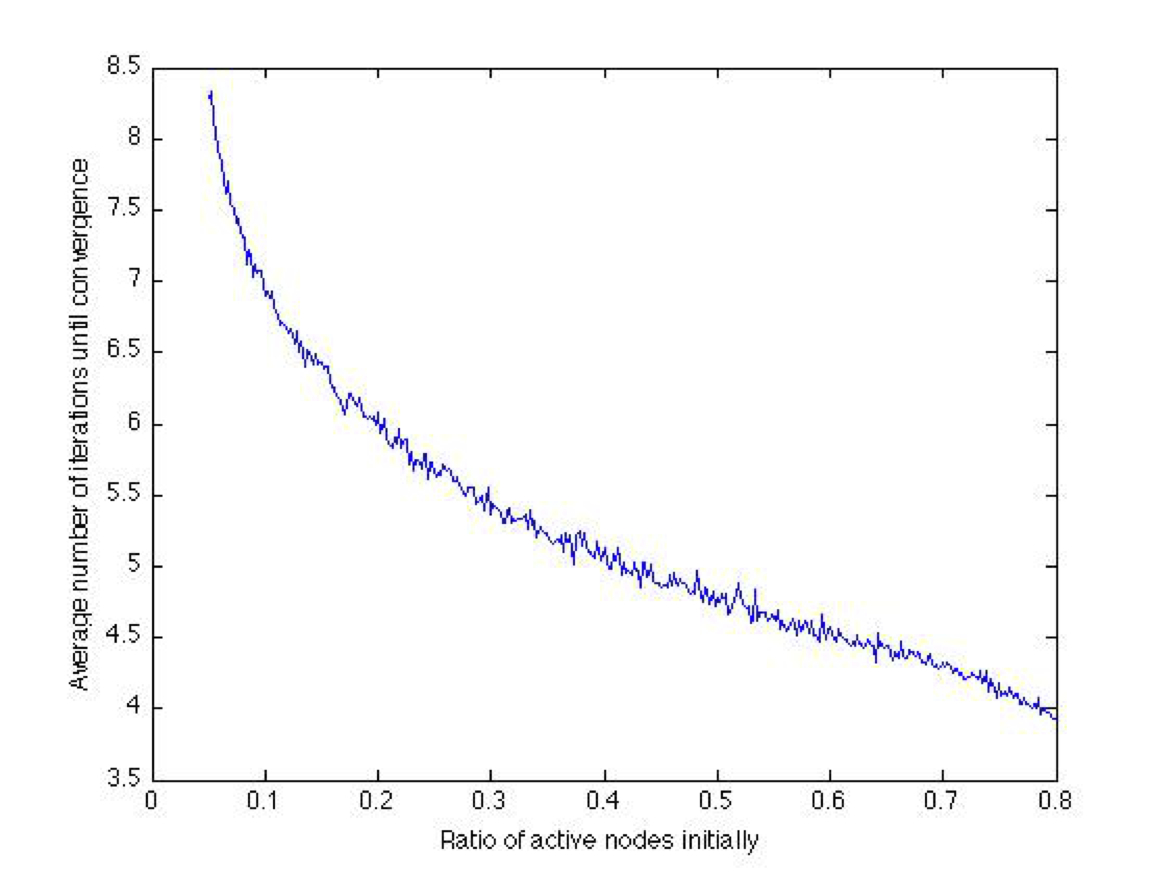}}
\subfloat[Threshold = 0.5.]{\label{fig:5}\includegraphics[width=0.33\textwidth]{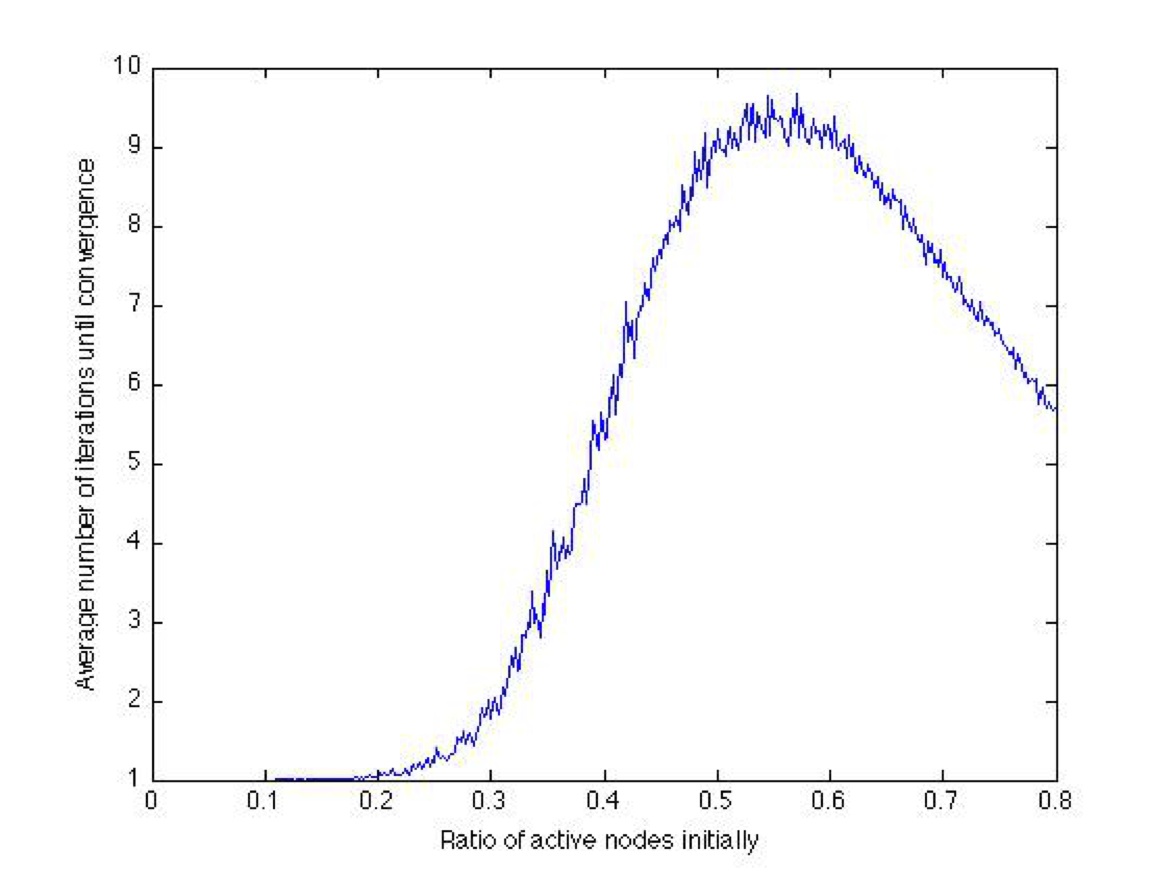}}
\subfloat[Threshold uniformly drawn from 0.05 to 0.8.]{\label{fig:6}\includegraphics[width=0.33\textwidth]{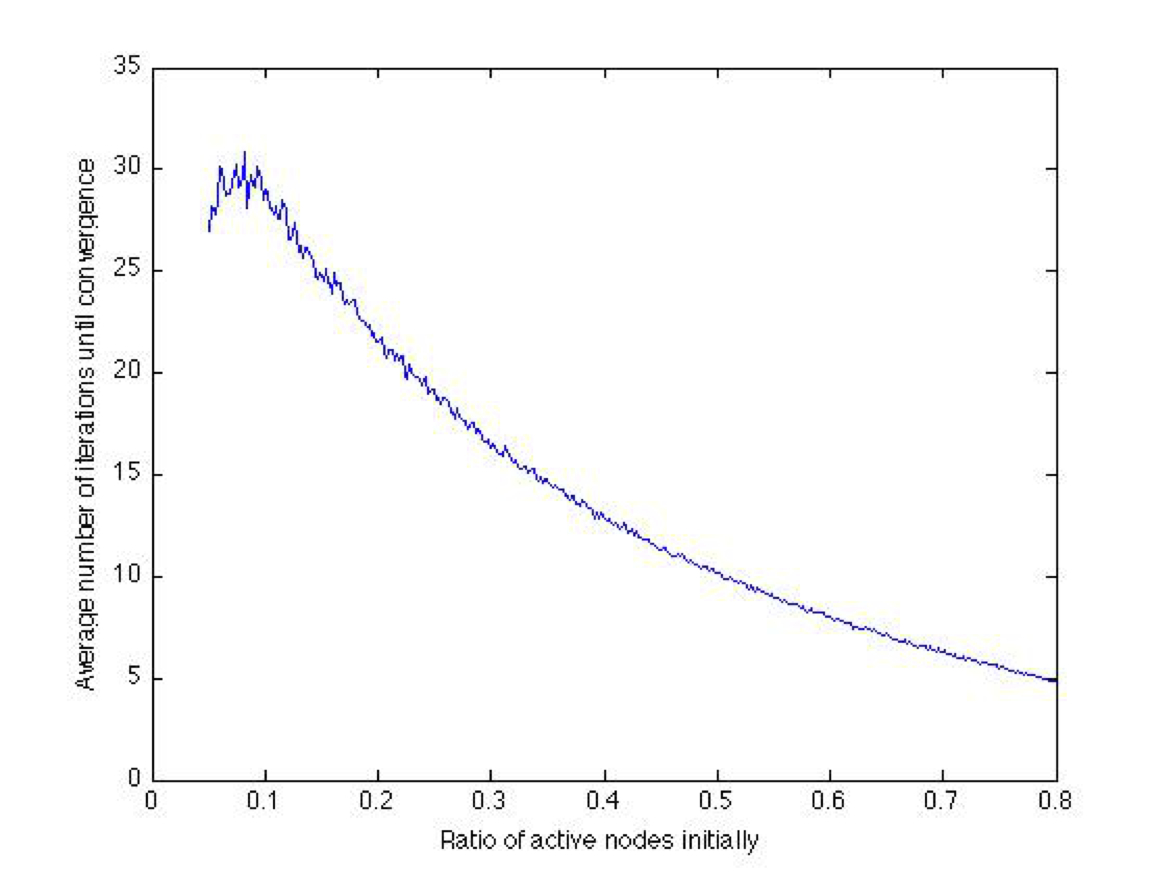}}

 \caption{Simulations for binary rating social contagion process in a small-world social network of 1000 nodes.}%The initially active nodes are in State "like" with probability 0.7 and in State "dislike" with probability 0.3. }
 \label{fig:fig}

\end{figure*}

We note that if assuming the time interval during which a user decides to buy the item recommended is short, then we only need to consider the influence of the user's immediate friends under a unit time step or two steps, which means the system only requires local social network information. 

We can view this as a networked cooperation game \cite{JonKleinberg}. Each node in the social network has three possible opinions on item $i$: \emph{like}, \emph{dislike}, and \emph{inactive}. Because of the homophily effect of social networks, if two nodes are connected in the social network, there is an incentive for them to have their opinions match.  We represent the payoff matrix as in Table I.

\begin{table}[htdp]
\begin{center}

\caption{Payoff for the three-strategy coordination game} 
\begin{tabular}{c r|c|c|c|}
\multicolumn{2}{c}{} & \multicolumn{3}{c}{$w$}\\ 
\multirow{1}{.05cm}{} 
 & \multicolumn{1}{c}{} & \multicolumn{1}{c}{\emph{like(1)}} & \multicolumn{1}{c}{\emph{dislike(-1)}} &\multicolumn{1}{c}{\emph{inactive(0)}}\\ \cline{3-5}
 \multirow{3}{.05cm}{$v$} 
 & \emph{1} & $a_{vw}, a_{wv}$ & $-a_{vw}, -a_{wv}$ & $-f(P_{v,i}^{cf},1), 0$ \\ \cline{3-5}
 & \emph{-1} & $-a_{vw}, -a_{wv}$ & $a_{vw}, a_{wv}$ & $-f(P_{v,i}^{cf},-1), 0$ \\ \cline{3-5}
 & \emph{0} & $0, -f(P_{w,i}^{cf},1)$ & $0, -f(P_{w,i}^{cf},-1)$ & $0,0$ \\\cline{3-5}
\end{tabular} 

\end{center}
\label{game}
\end{table}

Each node are playing many copies of this game with all its neighbors at the same time.  $a_{vw} > 0$ represents the payoff/penalty node $v$ receives if it coordinates/discoordinates with node $w$, and $f(P_{v,i}^{cf}, \cdot)>0$ represents the penalty for node $v$ playing active strategy if its neighbor is \emph{inactive}. The total payoff is the sum of all payoffs of individual games. Then the payoffs for node $v$ are:
\begin{itemize}
\item $\displaystyle\textrm{Payoff}_1 = \sum_{w\in N(v)}\left(a_{vw}S_{w,i} - f(P_{v,i}^{cf},1)\mathbf{1}_{\{S_{w,i}=0\}}  \right)$\\
\item $\displaystyle\textrm{Payoff}_{-1} = \sum_{w\in N(v)}\left(-a_{vw}S_{w,i} - f(P_{v,i}^{cf},-1)\mathbf{1}_{\{S_{w,i}=0\}}\right)  $\\
\item $\displaystyle\textrm{Payoff}_{0} = 0 $\\
\end{itemize}

If node $v$ chooses the strategy which provides maximum payoff and $f(P_{v,i}^{cf},\cdot)$ is reversely proportional to the number of inactive nodes, the solution matches our social contagion model.

\subsubsection{General Case}\label{subsec:general}
For a $1-$to$-R$ scale rating system, we pair up \emph{like}'s and \emph{dislike}'s in different levels. Denote $\mathcal{R} = \{1, 2, \cdots, R\}$ as the set of ratings and $\mathcal{S} = \{ \pm1, \pm2, \cdots, \pm\lfloor\frac{R}{2}\rfloor\}$ as the set of active state of different levels of \emph{like}'s and \emph{dislike}'s.  ``active-$0\textrm{''}\in \mathcal{S}$ if $R$ is odd. Define the mapping function $f_r:\mathcal{R} \rightarrow \mathcal{S}$
\begin{equation}
\label{ }
 f_r(r) = r-\bar{r},
\end{equation}
where 
$$
\bar{r}  = \left\{\begin{array}{ll}\lceil\frac{R}{2}\rceil  & \textrm{if }R \textrm{ is odd}\\ \frac{R}{2} & \textrm{if } R \textrm{ is even and } r > \frac{R}{2} \\\frac{R}{2}+1& \textrm{if } R \textrm{ is even and } r \le \frac{R}{2} \end{array}\right..
$$
E.g., for a 1-5 scale rating system, we assign ``-2'', ``-1'', ``active-0'', ``1'', ``2'' respectively on ratings 1 through 5. At each time step, let
\begin{equation}
\label{eq:multi }
S = \arg\max_{s\in \mathcal{S}}\left|\sum_{\substack{w\in N(v)\\|S_{w,i}| = s}}b_{vw} \cdot \textrm{sign}\left(S_{w,i}\right)\right|,
\end{equation}
% where $$f_s = \left\{\begin{array}{ll} 1 & \textrm{if }S_{w,i} = \textrm{active}-0\\ \textrm{sign}(S_{w,i}) & \textrm{otherwise}\end{array}\right.$$

\begin{equation}
\label{eq:sta }
%S_{v,i} = \left\{\begin{array}{ll} S\cdot \textrm{sign}\left(\sum_{\substack{w\in N(v)\\|S_{w,i}| = S}}b_{vw} \cdot S_{w,i}\right) & \textrm{if }|\sum_{\substack{w\in N(v)\\|S_{w,i}| = S}}b_{vw} \cdot \textrm{sign}\left(S_{w,i}\right)| \ge \theta_{v,i},\\ \textrm{inactive-}0 & \textrm{otherwise.}\end{array}\right.
S_{v,i} = \left\{\begin{array}{ll}\textrm{inactive-}0,  \textrm{if }|\sum_{\substack{w\in N(v)\\|S_{w,i}| = S}}b_{vw} \cdot \textrm{sign}\left(S_{w,i}\right)| < \theta_{v,i}\\ S\cdot \textrm{sign}\left(\sum_{\substack{w\in N(v)\\|S_{w,i}| = S}}b_{vw} \cdot S_{w,i}\right) ,  \quad\textrm{otherwise}\end{array}\right..
\end{equation}

$\theta_{v,i}$ in Equation \ref{eq:sta } is the influential threshold.
%\begin{equation}
%\label{ }
%\theta_{v,i}  = \left\{\begin{array}{ll}\theta & \textrm{if }f_r\left(P_{v,i}^{cf}\right)\cdot\left(\sum_{\substack{w\in N(v)\\|S_{w,i}| = S}}b_{vw} \cdot S_{w,i}\right) < 0  \\ \min\{\theta, 1- \theta \} & \textrm{otherwise}\end{array}\right.,
%\end{equation}
%where $\theta = f_r\left(P_{v,i}^{cf}\right)/\lfloor\frac{R}{2}\rfloor$.

Here we assume that the mean-state ``active-0'' does not provide valuable information thus it cannot influence other nodes. We can easily change this assumption by making a minor modification in Equation \ref{eq:multi } and \ref{eq:sta }. The social influential prediction on a target user $u$ is

%In reality, strong opinions on a certain item provide more information thus  is more likely to spread over the network, we can simplify the model by only taking extreme ratings (highest and lowest levels) into account. In that case, the model is then similar as in Section \ref{subsec:simple}.

\begin{equation}
\label{ }
P_{u,i}^{si} = \left\{\begin{array}{ll} f^{-1}(S_{u,i}) & \textrm{if }u \textrm{ is activated}  \\ \textrm{inactive-}0 & \textrm{otherwise}\end{array}\right..
\end{equation}

\subsubsection{Simulation Results}
We simulate the social contagion process of binary rating model as shown in Fig.\ref{fig:fig}. We construct a social network generated by Watts and Strogatz model \cite{Watts} of 1000 nodes, and assign different thresholds for the nodes and analyze the convergent rate and speed. Initially, active nodes are in State \emph{like} with probability 0.7 and \emph{dislike} with probability 0.3. We assign influential factors by a uniform random partition of a unit influence among neighbors of a node. Results are the average of 500 independent simulations. When the threshold is low, as in Fig.\ref{fig:1} and Fig.\ref{fig:4}, even a very small portion of initial active nodes can activate almost all the inactive nodes in the network, into both majority state \emph{like} and minority state \emph{dislike}. With the increasing number of the initial active nodes, the newly activated nodes dominatingly choose the majority state and the number of iterations to convergence decreases. This shows that in a susceptible social network, social contagion is likely to progressively influence all the nodes. When we increase threshold to 0.5, as shown in Fig.\ref{fig:2} and Fig.\ref{fig:5}, social contagion process converges faster, but small ratio of initial active nodes cannot influence the network. Fig.\ref{fig:3} and Fig.\ref{fig:6} show a more realistic setting of the threshold: each node's threshold is drawn randomly from a uniform distribution from 0.05 to 0.8, representing the various susceptibility of different people. When 20\% of nodes are initially activated, more than half of the unactivated nodes will become active to the majority states eventually, this rate stays stable when the ratio of initial active nodes increases, while the convergence speed increases.  Our simulations show that in a general setting, the result of social contagion process is users' local estimation of majority opinions with the penalty from the disagreement with their own estimation. This also follows our coordination game explanation. For more general case, we expect that the ratings of newly activated nodes to be clustered according to their social network structure, generally following the majority opinion.

\subsection{Discussions on Recommendation to Individuals}
Define susceptibility factor $\alpha_u \in [0,1]$, which is the attribute of a user.  The more susceptible the user is, the higher the value of $\alpha_u$ should be. The recommendation prediction  $P_{u,i}$ to a target user $u$ on a certain item $i$ is calculated by 
\begin{equation}
\label{recom}
P_{u,i} = \left\{\begin{array}{ll} P_{u,i}^{cf} & \textrm{if }P_{u,i}^{si} = \textrm{inactive-}0 \\(1-\alpha_u)P_{u,i}^{cf} + \alpha_u P_{u,i}^{si}& \textrm{otherwise}\end{array}\right.
\end{equation}

Then we can recommend the top-$k$ predicted items to the target user. In particular, if we set $\alpha_u = 1$ for $\forall u$, Equation \ref{recom} becomes
\begin{equation}
\label{ }
P_{u,i} = \left\{\begin{array}{ll} P_{u,i}^{cf} & \textrm{if }P_{u,i}^{si} = \textrm{inactive-}0 \\ P_{u,i}^{si}& \textrm{otherwise}\end{array}\right.
\end{equation}
If we ignore the inactive ratings, the recommendation model above becomes a recommender based on ratings by users' immediate and distant friends. It can be a supplement of the existing recommendation systems and provide users extra information of the opinions from someone they would trust.

\section{Recommendations for Groups}\label{sec: groups}

To show how our group recommendation model differs from aggregating each user's predictions or merging preference profiles,  let us begin with a brief example: Jessica and Mike want to see a movie with Eric, a friend visiting them. When choosing the movie to watch, the couple would like to follow their guest's opinion. Eric, accommodating as he is, also tries to take Jessica and Mike's taste into consideration. After some discussion, they finally agree on a movie that Eric has wanted to see for a long time, and also is interesting to both Jessica and Mike.  We will now introduce a \emph{social influence model} to describe Jessica, Mike and Eric's opinion evolution in their discussion.

\subsection{Social Influence Model}\label{subsec:group model}
\emph{Social influence theory} \cite{Friedkin}\cite{Hui} describes how a network of interpersonal influence enters into the process of opinion formation, which postulates a recursive definition for the interpersonal influence  in a group of $N$ users:
\begin{equation}
\label{eq:8}
P_i^{(t)} = AWP_i^{(t-1)}+(I-A)P_i^{(1)},
\end{equation}
for $t = 2, 3,\ldots$.  $P_i^{(1)}$ is an $N \times 1$ vector of users'  initial opinions on an item $i$, it can be either existing individual rating $r_{u,i}$  or preference prediction $P_{u,i}$, or a mixture of both. $P_i^{(t)}$ is an $N \times 1$ vector of users' opinions at time $t$.  $W = [w_{uv}]$ is an $N \times N$ matrix of interpersonal influences with $0\le w_{uv} \le 1, \sum_j^Nw_{uv}=1$, and $A = \textrm{diag}(\alpha_{11}, \alpha_{22}, ..., \alpha_{NN})$ is an $N \times N$ diagonal matrix of users' susceptibilities to interpersonal influence on item $i$ with $0 \le \alpha_{uu} \le 1$.  This susceptibility factor is a user's attribute and by default it is associated with $\alpha_u$ in Equation \ref{recom}, but can also be set by users. When $A = 0$, meaning no group members are willing to change their own opinions in response to other group members' tastes. In that case, $P_i^{(t)} = P_i^{(1)}$. This is normally what a traditional group recommender assumes. However, although individuals in a group may not necessarily reach consensus, they tend to coordinate their decisions through the effect of social influence \cite{Cosley}. This is the initial motivation that we apply \emph{social influence network theory} to recommendation model for groups. 

When we consider the process in an equilibrium state (assuming convergence), Equation \ref{eq:8} becomes
\begin{equation}
\label{ }
P_i^{(\infty)} = AWP_i^{(\infty)} + (I-A)P_i^{(1)}.
\end{equation}
If $I-AW$ is nonsingular, then
\begin{equation}
\label{eq:inf}
P_i^{(\infty)} = VP_i^{(1)},
\end{equation}
where
\begin{equation}
\label{eq:v}
V = (I-AW)^{-1}(I-A).
\end{equation}

$V$ is a matrix describing the total interpersonal influence that affect a user's evolution on a certain item. $P_i^{(\infty)}$ is an $N \times 1$ vector describing users' final preference prediction on a certain item $i$ under social influence.  This can be viewed as a compromised disagreement after sufficient consideration on $N-1$ other group members' opinions.  Then we can follow the normal regime, such as average rating, to aggregate users' opinions to form a recommendation list for the group.

For example, in the scenario at the beginning of this section, Jessica, Mike and Eric may set $A = \textrm{diag}(0.9, 0.9, 0.5)$ and $W =[\textrm{0.1 0.1 0.8; 0.1 0.1 0.8; 0.25 0.25 0.5}]$. For movie $A$, suppose their individual predicted ratings are $P_A^{(1)} = (2, 3, 5)$. In above setting, by Equation \ref{eq:inf} and \ref{eq:v}, we have $P_A^{(\infty)} = (4.52, 4.62, 4.86) $. Now the average rating of $A$ for the group is 4.66 instead of 3.33 if averaging directly from $P_A^{(1)}$. We can see that the result is inclined to Eric as it initially assumed.

\subsection{Discussions on Possible Applications for Group Recommendation}
The recommendation model proposed in Section \ref{subsec:group model} can be applied and added to various recommendation systems with different settings of interpersonal influence factor $W$ and susceptibility factor $A$. For instance, for restaurant recommendation, it is likely that group members are willing to go to a good restaurant even though some of them have already dined there before. Thus original opinions $P_i^{(1)}$ in Section \ref{subsec:group model} can be a combination of existing ratings and predicted ratings, and with a higher susceptibility for the latter. On the other hand, for trip recommendation and movie recommendation, group members are likely to explore something unknown to everybody, thus $P_i^{(1)}$ are set as original predictions of individual preference. In both cases, the recommendation model provides a way to take interpersonal influence within a group into account before preference aggregation procedure.

\subsection{Social Contagion Model vs. Social Influence Model}  Recommendation to individuals provides items that a target user is most likely to be interested in.  \emph{Social contagion model} in Section \ref{sec: individuals} is a probabilistic framework that simulates how an opinion on a certain item spreads through the social network.  However, the buying decision made by the user is a personal choice: the user may listen to others' opinions but there will be no compromise in the final decision.  In addition, the active nodes at the beginning are users who have already made their explicit or implicit statement of preferences and will not change with time. Since the time for the target user to decide whether to buy the recommended item is usually short, it is reasonable to model the social influence to individual users as a progressive process. On the other hand, a single decision for a group of people is not only based on fairness, as in most of the current group recommenders, but also based on group members' susceptibility and influence power. We assume group members will or are willing to modify their choices according to other group members' opinions, and it is usually the case in social activities.  \emph{Social influence model} shows us the compromised disagreement formed iteratively by interpersonal influence. 

\section{Conclusions and future work}
\label{sec: conclusions}

In this paper, we propose two social influential recommendation models, for individuals and for groups respectively. The individual recommendation is based on \emph{social contagion}, which  takes the effect of homophily among friends and effect of word of mouth into consideration while making recommendations. In the recommendation model for groups, we provide preference specifications that not only reflect subjective personal taste but also opinion evolution with the knowledge of other members' preferences. 

For future work, we plan to validate our models through a real online social network dataset from Yelp.com, which provides users' ratings of restaurants, spas, etc. Yelp  also provides social network feature which is ideal to compare our model with the performance of collaborative filtering recommendation and some existing recommendation systems based on social networks such as SNRS \cite{He}.  We also want to use machine learning techniques to cluster different types (e.g. subjective, susceptible, hybrid, etc. ) of users,  and explore the well-performing settings for each type on  threshold $\theta_{u,i}$, susceptibility $A$ and interpersonal influence weight $W$. Furthermore, we want to apply our group recommendation model in different application areas such as movie recommendation, restaurant application, etc.

\bibliographystyle{plain}

\end{document}